\documentclass[11pt]{amsart}

\usepackage{amsmath}
\usepackage{amssymb}
\usepackage{graphicx}

\newtheorem{theorem}{Theorem}[section]

\theoremstyle{definition}

\newcommand{\eps}{\varepsilon}

\numberwithin{equation}{section}

\title[Lax Pairs: Integrable, Less Integrable and Nonintegrable Systems]{Lax Pairs: Integrable, Less Integrable and Nonintegrable Systems}

\author[D.~C. Antonopoulou]{D.~C. Antonopoulou}

\author[S. Kamvissis]{S. Kamvissis}

\thanks{$^{\dag}$ Research funded by ARISTEIA II grant no. 3964 from
the General Secretariat of Research and Technology, Greece}

\address[D.~C. Antonopoulou]{Department of Mathematics, National and Kapodistrian University
of Athens, Greece} \email{\tt danton@math.uoa.gr}

\address[S. Kamvissis]{Mathematics Building, University of Crete, GR--700 13 Voutes, Greece,
and, Institute of Applied and Computational Mathematics, FORTH,
GR--711 10 Voutes, Greece} \email{\tt spyros@tem.uoc.gr}

\keywords{Complete integrability of PDEs, Lax Pair, Long Time
Asymptotics,  Initial-Boundary Value PDE problems, Numerical
experiments.}

\subjclass[2020]{37K10, 37K40, 35Q51, 65M06}

\begin{document}

\begin{abstract}
Completely integrable finite dimensional Hamiltonian systems are
well understood thanks to the work of Liouville and Arnold. On
the other hand, the Lax Pair formulation of the KdV equation
marks the beginning of the extension of the completely integrable
theory to infinite dimensional Hamiltonian systems. Solutions of
initial value problems for systems that admit a Lax Pair
formulation normally have a tame qualitative behavior if Lax
Pairs  give rise to an infinite complete set  of conserved laws.
The situation is different for initial-boundary value problems,
even in one space dimension. There are problems where
integrability persists and regular (long time asymptotic)
behavior can be proven. There are others where even irregular
"fractal-chaotic-looking" behavior can appear. In this short
article we review an instance of each case. We also make a
connection with results  from the existing theory of perturbed
Lax Pair equations on the real line.
\end{abstract}

\maketitle
%
%
%
$ To ~~the ~~memory ~~of ~~Peter ~~Lax, ~~teacher ~~and ~~mentor$

\bigskip

\bigskip
\section{Introduction}

The Lax Pair formulation of the KdV equation \cite{L} marks the beginning of the extension of the theory of completely integrable systems
to Hamiltonian systems of infinite dimension. Initial value problems for systems admitting a Lax Pair formulation are well understood if the initial data satisfy some
decay or convergence conditions at infinity or are  periodic, enabling the possibility of  solution via inverse methods (inverse scattering or inverse spectral algebro-geometric method).
A crucial fact is that Lax Pairs give rise to an infinity of conservation laws. Ensuring appropriate initial data so that the conserved quantities are finite
and reducing the system appropriately so that a complete set of  action and angle variables exists, the system is completely integrable in the sense that the solution
is reducible to the solution of a (local) Riemann-Hilbert factorization problem (in the case of one space dimension) or a nonlocal Riemann-Hilbert factorization problem
or a $\bar \partial$-problem (in higher space dimensions). Such problems are generally posed on a Riemann surface and are amenable to asymptotic analysis.
For local Riemann-Hilbert factorization  problems one applies the so-called nonlinear stationary phase and steepest descent methods;
there is already a huge literature, see e.g. \cite{DZ}, \cite{KMM}, \cite{KT}. For $\bar \partial$-problems there are fewer results (e.g. Perry \cite{P});
for nonlocal Riemann-Hilbert factorization problems see Donmazov, Liu and Perry \cite{DLP}.

Things are different  for initial-boundary value problems, even in one space dimension.
There is an extension of the inverse method (a "unified transform method", developed by Fokas and his
collaborators) \cite{F},  \cite{F2}, \cite{F3}, \cite{FIS}.
But a crucial feature of this method is that it
requires the values of more  boundary data than given
for a well-posed problem. This has two important consequences. First, it somehow lowers the degree of effectiveness of the asymptotic formulae
since they involve knowledge of scattering data associated  to values of quantities not given; they are only very implicitly given as limits of $x-$derivatives of the solution near $x=0$.
This is not so unfortunate if we can prove that the map that takes the given data to the unknown (but necessary) values (a generalised Dirichlet-to-Neumann map) is continuous.

Most crucially however, for the unified transform theory to apply, all (overdetermined) data given have to be in an appropriate class (e.g. decaying or periodic + decaying). But it is not at all clear that if the data required for a well-posed problem are such, then the extra (superfluous, implicitly defined) "data" are also in an appropriate class.
This is a problem one must address before going on.

More specifically, in the case of cubic NLS knowledge of the Dirichlet data suffices to
make the problem well-posed but the unified transform method also requires
knowledge of the  values of the Neumann "data". The study of the Dirichlet-to-Neumann map
is thus  necessary $before$ the application of the unified
transform. In the papers \cite{AK1}, \cite{AK2}, we presented  a rigorous study of this map
for a large class of decaying Dirichlet data. We showed that the values of the
Neumann "data" are also sufficiently decaying and that the Dirichlet-to-Neumann map defined in the relevant function spaces is $continuous$; hence that  the unified transform
method $can$ be applied.  These results are presented in the next section.

In the third  section we present the long time asymptotic formula for the defocusing NLS problem, which can be derived from the Riemann-Hilbert formulation, which in view of the result of section 2 is now rigorously justified. 

In the fourth  section we present some beautiful numerical
experiments by Arthur, Dorey and Parini, which clearly show the
existence of irregularities in the behavior of the Sine-Gordon
initial-boundary value problem with initial  data and a Robin
boundary condition. Furthermore the Dirichlet boundary function
$u(x,0)$  seems to be  unbounded. This is a clear instance of  a
problem which admits a Lax pair formulation but is  not
integrable: adding a boundary and  boundary  conditions, even
ensuring a uniquely solvable problem may or may  not preserve
integrability! \footnote{This phenomenon should not be confused
with the so-called deterministic turbulence \cite{L2}, \cite{BGK}
appearing in initial value problems that $can$ be studied via the
inverse scattering transform and Riemann-Hilbert problems
\cite{KMM}. Deterministic turbulence is an integrable phenomenon.}

In the fifth section we compare  our results on initial-boundary
problems to existing results on initial value problems for
$perturbations$ of Lax equations. In the sixth section we present
some numerics for a problem where we have not been able to prove so far
that the Neumann values are  appropriately decaying. Still the
results are what one would expect if they did.

\section{NLS}

Consider the NLS  equation with cubic non-linearity,  posed on the real
positive semi-axis $\mathbb{R}^+$
\begin{equation}\label{nls}
{\rm i}q_t+q_{xx}-2\lambda|q|^2q=0,\;\;\;\;x> 0,\;\;\;\;0<
t< +\infty,
\end{equation}
and initial-boundary data
\begin{equation}\label{ic}
\begin{split}
&q(x,0)=q_0(x),\;\;\;\;0\leq x<+\infty\\
&q(0,t)=Q(t),\;\;\;\;0\leq t<+\infty,\\
\end{split}
\end{equation}
where $q_0, Q$ are classical functions satisfying the
compatibility condition $q_0(0)=Q(0)$.

The case $\lambda =1$ is the defocusing case, while $\lambda =-1$ is the focusing case.

Back in 1991, Carrol and Bu in \cite{CarBu} established the existence of a
unique global classical solution $q\in C^1(L^2)\cap C^0(H^2)$ of
the problem \eqref{nls}-\eqref{ic}, with $q_0\in H^2$, $Q\in
C^2$ and $q_0(0)=Q(0)$, by using PDE theory. Later papers like \cite{Hol} by Holmer etc. have also provided results in Sobolev cases.
For our purposes the classical result in \cite{CarBu} suffices.

On the other hand, it is well-known \cite{ZaSh} that the non-linear  Schr\"odinger equation
(NLS) with cubic non-linearity can be written as a Lax pair and that, at least the Cauchy problem
is `completely integrable'; this means that there is an infinity of conservation laws which are in Poisson involution, and furthermore that
the problem can be linearised via the scattering transform. It does not mean that there is a bona fide explicit solution; at best the inverse scattering problem,
rewritten as a Riemann-Hilbert factorization problem can be effectively treated asymptotically. $Effective$ long time, long range and semiclassical asymptotic formulas can be provided:
 they depend on the initial data either very explicitly or at worst via the solution of simple linear ODEs.

In \cite{FIS} the authors use the unified transform method to solve the  problem on the real
positive semi-axis, given values for the initial data and Dirichlet data
(which render the problem well-posed) and also the values of the $Neumann$ "data" $P(t):=q_x(0,t)$. What is required for that
theory to work (for infinite time) is that the  Neumann function $P$ (as well as the
Dirichlet data) lives in some class with nice decaying properties
such that the  unified scattering transform can be properly defined.
This is exactly the content of our theorems below: we provide
several reasonably inclusive large classes of Dirichlet data,
such that both Dirichlet and Neumann functions  decay as
$t\rightarrow\infty$ fast enough for the scattering method to
work. Hence \cite{FIS} applies,  a Riemann-Hilbert factorization problem is possible,
and explicit asymptotics (long time \cite{FIS}, long space, or even semiclassical \cite{K} \cite{FK}) are available.
These formulae are not as effective as the formulae for the Cauchy problem. The reason is that in general the Dirichlet to Neumann
map is very implicit. So some functions appearing in the asymptotic formulae involve scattering data related to the Neumann boundary values;
these $cannot$ be effectively computed.
Still the Dirichlet-to-Neumann map $is$ continuous here; in later sections we consider more complicated problems
where the dependence can be very unstable.

Our main result concerning the defocusing case is the following, see \cite{AK2}.

\begin{theorem}
Let $q$ be the
unique global classical solution $q\in C^1(L^2)\cap C^0(H^2)$ of
the initial-value problem for defocusing NLS, with  Dirichlet data $Q\in C^2$ and
$Q(0)=q_0(0)$.

Assume that $q_0\in H^1(0,\infty)\cap L^4(0,\infty)$ and $x
q_0\in L^2(0,\infty)$.

If $q(0,t)$, $q_t(0,t)$ have a sufficiently fast decay as
$t\rightarrow\infty$, that is  $\mathcal{O}(t^{-\alpha})$ and $\mathcal{O}(t^{-\beta})$, for $\alpha>
3/2$ and $\beta> 5/2$ respectively, then
$$\int_0^\infty |q_x(0,t)|dt<\infty.$$

Furthermore, if  the Dirichlet data  belong in the Schwartz class, then the
Neumann data  also belong in the Schwartz class. The Dirichlet-to-Neumann map is continuous in the appropriate $infinite ~time$ spaces.
\end{theorem}

As mentioned, this implies that a Riemann-Hilbert factorization problem is possible,
and explicit long time asymptotics  are available
\footnote{This is actually what we prefer to call  "integrability".
It presupposes the existence of Lax pairs $and$ a useful inverse theory!}.
In the next section we present the main long time asymptotic formula for the defocusing case.

We also have a result for the focusing case. Here, we have to assume that some data are small.

Let $q$ be the
unique global classical solution $q\in C^1(L^2)\cap C^0(H^2)$ of
the initial-value problem for focusing NLS, with  Dirichlet data $Q\in C^2$ and
$Q(0)=0$. Assume for simplicity that the initial data is zero.

Also, let
$$\int_0^\infty|Q(t)|^2dt,$$
be sufficiently small.  If as $t\rightarrow\infty$
$$Q(t)=\mathcal{O}(t^{-5/2-\eps}),\;\;\;\;Q_t(t)=\mathcal{O}(t^{-5/2-\eps}),
\;\;\;\;Q_{tt}(t)=\mathcal{O}(t^{-1/2-\eps}),$$ for some small
$\eps>0$, then there exists $c>0$ independent of $t$ such that for
 $t$ large
\begin{equation}
\int_0^\infty|q_{x}(0,t)|dt<\infty.
\end{equation}
Furthermore, if the Dirichlet data  belong in the Schwartz class, then the
Neumann data  also belong in the Schwartz class.

\section{Long time asymptotics}

From the Riemann-Hilbert formulation one can derive precise long
time asymptotics. For defocusing NLS this was first done in
\cite{DIZ}. Their calculation was for the initial value problem.
However, since the Riemann-Hilbert problem for the
initial-boundary value problem is actually very similar,  the
same  computation gives rise to the following long time
asymptotics, as cited in \cite{FIS}

\begin{equation}
q(x,t)=  \frac{a(\frac{-x}{4t})}{ t^{1/2} } e^{ ix^2/4t  +2i   (a(\frac{-x}{4t}) )^2 log t + i \phi (\frac{-x}{4t} ) } (1+o(1)) ,
\end{equation}
as $t \to \infty$,
where the functions $a, \phi$ are given by simple explicit formulae  depending on the scattering data corresponding to the Dirichlet and Neumann data.
We repeat that only the Dirichlet data are given (for a well-posed problem) and the Neumann values are only implicitly determined from the initial and Dirichlet data.
Because of the continuity of Dirichlet-to-Neumann map \cite{AK2} and  continuity properties of the scattering map \cite{Z}  the functions $a, \phi$ depend continuously on the data.

These asymptotics are uniformly valid in any closed linear sector (with half-lines as boundaries) that lies entirely in the open first quadrant $x,t >0$.
The same expression gives the long range asymptotics as $x \to \infty$\footnote{with o(1) error  as $x \to \infty$}.

Near the boundary $x=0$ the asymptotic analysis is more delicate and depends on the details of the behavior of the Dirichlet data as
$t \to \infty$ \footnote{A careful demonstration has been presented in \cite{AL} for the derivative NLS equation in the case of Schwartz data.}. If the Dirichlet data are
unbounded the (currently existing) uniform transform method  is not  applicable.

A very similar asymptotic formula also holds in the focusing case if no solitons are present
(which is shown to be true in \cite{AK1} for zero initial data and small Dirichlet data).
However, in general, and assuming that the unified theory applies,  one would have to add a  sum of soliton terms.
For large times these solitons would separate, the tallest being also  the fastest. So, for a bounded set (above and below by positive constants) of values of $x/t$
(corresponding to the soliton velocities) the leading asymptotic term is given by a
1-soliton formula with  parameters depending on the scattering coefficients corresponding to the given (initial and  Dirichlet) data $and$ also the Neumann values
or just a decaying term!
For the actual details, see appendix B of \cite{FIS}. Of course if we don't know how to control the Neumann values, and so far we do not, nothing is rigorously proven.
At best we can provide some convincing numerics that this is the case: the asymptotics are given by a finite set of solitons travelling to the right and a decay term.
This is what we do in section 6.

We conclude this section with a short comment on the periodic
case: consider focusing NLS with decaying initial data and
periodic Dirichlet data. It is an open question whether the
solution is asymptotically periodic for large times. Again a
crucial ingredient would be provided by information on the
asymptotic periodicity of the Neumann values. We refer to
\cite{BKSZ} for some theoretical analysis and some numerics for
focusing \footnote{see also \cite{Le} for the defocusing case}
NLS, which suggest  "integrability" in some \footnote{genus zero}
cases, but the  problem remains  open in general\footnote{In the
case of genus zero, the numerics suggest that the Neumann
function can be of higher genus; the general problem may then be
terribly complicated even for finite genus, since this raises the
possibility that Dirichlet and Neumann functions have a different
algebro-geometric structure in general. At the level of rigorous
justification a uniform control of the scattering data involved
has eluded us so far even in the simplest cases.} \footnote{Some
attempts at a formal level have been made in \cite{FL},
\cite{BL}, \cite{LF}. In our opinion they just underscore the complexity of the
intractable original problem!}.

\section{Sine-Gordon with Robin condition}

The results in the previous section could lead one to suspect that bona-fide integrability could hold for all sorts of initial-boundary value problems. On the other hand that would be really surprising in view of the fact that an infinite set of conserved quantities cannot in general survive.

Following \cite{ADP}
we consider the equation
\begin{equation}
u_{tt}-u_{xx}+sinu=0
\end{equation}
in the half plane $x<0$ with a homogeneous Robin condition
\begin{equation}
u_x + 2ku=0
\end{equation}
at $x=0$.
Here $k$ is a given real constant. This is an intermediate condition between a Neumann and a Dirichlet condition.
For $k=0$ we recover a Neumann condition and for  infinite $k$  we recover a Dirichlet condition at $x=0$.

Consider initial data of one-antikink form: $$ u(x,t) = 4 arctan
(e^{-\gamma (v_0) (x-v_0 t- x_0)}),$$ that is take $ u(x,0),
u_t(x,0)$ corresponding to the one-antikink formula. Note that
the one-antikink  is localised near $x_0<0$ so initially the
effect of the boundary is not felt (much), if it is chosen far
enough from the  boundary $x=0$.  Here  $\gamma (v_0) =
(1-v_0^2)^{-1/2},$ $v_0>0$  is the intial velocity \footnote{This
is expected (but never actually rigorously proven as far as we
know) to be a uniquely solvable initial-boundary value problem.
Of course the numerics of \cite{ADP} support this fact.}.
Numerical experiments in \cite{ADP} (Figures 8 and  14 in
particular) consider the boundary-initial value problem and focus
on the recovery of the Dirichlet  boundary value $u(0,t)$. There
are very careful  plots of $u(0,t)$ in terms of $v_0$ and $k$
with the choice $x_0=-30$ (that is away enough from the boundary).
The Dirichlet values (or alternatively the Neumann values) data
are necessary for the application of the unified transform even
though they are not part of the conditions defining the
well-posed problem; they are only implicitly defined  by the
solution itself. So the understanding of their behavior is
crucial for the applicability of the unified transform.

The result of the reflected wave at the boundary $x=0$ has to consist of breathers,
kinks  and  antikinks (and some decaying "background" term).
The intuitive non-rigorous reason for this is that away from the boundary one expects the effect of the initial data to be dominant,
because of the expected finite propagation speed (up to smaller error terms). \footnote{Of course the finiteness of the propagation speed can only be proved for initial value problems, using the Riemann-Hilbert formulation. It may only be proved for initial-boundary value problems if one knows already that the unified transform is applicable.}
But it is known that any initial data will give rise to a set of kinks, antikinks and breathers\footnote{this is the "soliton resolution" for Sine-Gordon}.

The most striking behavior of $u(0,t)$  is observed
for values of the real constant $k$ between $0.05$ and $0.07$
and the initial velocity $v_0$ between 0.875 and 0.9 and for large times  $t_f=x_0/v_0 +1000$.
The authors  observe the existence of  breathers and possibly an antikink.
Clearly unstable  phenomena occur: very slight changes in the  parameters can affect
(in a $fractal-chaotic-$looking way) the production (or not) of a reflected antikink!

Citing \cite{ADP}
"The dark blue bands, where $u(0, t )$ is near zero, correspond
to an antikink being emitted, while in the light green areas, where $u(0, t )$ is near $2\pi$,
only breathers are emitted. In between these areas are indeterminate regions where a
very slight change in the initial parameters can cause an antikink to be produced or
not. The oscillations in the boundary value of the field on the left of the plot are due
to a breather becoming trapped at the boundary, only decaying very slowly there, in
contrast to behaviour on the bottom right where this breather is able to escape and the
field relaxes to zero much more quickly."

Furthermore, the function $u(0,t)$ looks  $unbounded$ for large time.
When $k<0$ the unboundedness becomes more evident. Large $|k|$ allow for an infinite number of kinks accumulating at $x=0$.

This means that the map that takes the initial data $u(x,0), u_t(x,0)$ to the Dirichlet values $u(0,t)$ depends in a very unstable  way on $k$ and $v_0$.
This fact does not exclude the possibility
of applying the unified transform method.  What $does$ render the method inapplicable is the unboundedness of $u(0,t)$.
So we have two observations here: instability  $and$ inapplicability of the unified method, even though the data are Schwartz and the Robin condition is linear and
very innocent looking. The problem is non-integrable!

\section{Comparison with the perturbed NLS on the real line}

In \cite{DZ2} the authors consider the initial value problem for the defocusing NLS with an extra  perturbation term $\epsilon |u|^l u, ~~l>2, \epsilon>0$
and initial data decaying at infinity. What they discover is that for small  $\epsilon$ the problem is still integrable! In particular they derive long term asymptotics
similar to those in the unperturbed case.

Now, in a sense the initial-boundary value problem is a forced perturbation of the initial value problem\footnote{In fact,
the original PDE methods paper \cite{CK} on the initial-value problem for KdV treats it by comparing to a problem on the line with a delta-type extra force.}.
So it makes sense to compare the results of \cite{DZ2}
with our results in \cite{AK1} and \cite{AK2}. But it is also interesting that our "forcing" term is  not small in the defocusing case;
it has to be small enough only in the focusing case.

Also, no chaotic-looking phenomena are known in the fully non-integrable case of the perturbed NLS on the real line with large positive $\epsilon$. So it seems that the
initial-boundary value problem expresses a richer set of phenomena!

\section{Numerical approximation of the focusing NLS on the half line}
In this section we consider the focusing NLS equation on the half
line in the form
$$iq_t+\frac{1}{2}q_{xx}+|q|^2q=0,\;\;x\geq 0,\;\;t\geq 0,$$
with initial data
$$q(x,0)=q_0(x),\;\;\;q(0,t)=Q(t),\;\;x\geq 0,\;\;t\geq 0,$$
with $$q_0(x)\rightarrow 0\;\;\mbox{as}\;x\rightarrow\infty.$$ We
will also impose a decay condition at infinity given by
$$q(x,t)\rightarrow 0\;\;\mbox{as}\;x\rightarrow\infty.$$

We have approximated numerically the solution by applying a
nonlinear Crank-Nicolson finite differences scheme.

In more detail, for $x_{\rm fin}$ large enough, we consider a
uniform partition of $[0,x_{\rm fin}]$ (space)
$0=x_0<x_1<\cdots<x_J<x_{J+1}=x_{\rm fin}$ with step-length
$h=\frac{x_{\rm fin}}{J+1}$, and a uniform partition in time
$0=t^0<t^1<\cdots<t^N=T$ of $[0,T]$ with step-length
$k=\frac{T}{N}$.

The decay of $q$ as $x\rightarrow\infty$ is modeled by imposing
on the bounded interval $[0,x_{\rm fin}]$ the b.c. $q(x,t)=0$ at
$x=x_{\rm fin}$.

\subsection{The numerical scheme}
We seek
$$Q^n:=(Q_0^n,Q_1^n,Q_2^n,\cdots,Q_J^n,Q_{J+1}^n)\in\mathbb{C}^{J+2},$$ for
$n=0,\cdots,N$, with $$Q_j^n\simeq q(x_j,t^n),$$ for
$j=0,\cdots,J+1$, $n=0,\cdots,N$, where
$$Q_0^n:=q(0,t^n),\;n=0,\cdots,N,\;\;Q_j^0:=q(x_j,0)=q_0(x_j),\;j=0,\cdots,J+1,$$
$$Q_{J+1}^n:=0,\;n=0,\cdots,N,$$ satisfying
\begin{equation}\label{scheme}
\begin{split}
i\frac{Q_j^{n+1}-Q_j^{n}}{k}+\frac{1}{2}\frac{Q_{j-1}^{n+1/2}-2Q_{j}^{n+1/2}+Q_{j+1}^{n+1/2}}{h^2}+|Q_j^{n+1/2}|^2Q_j^{n+1/2}=0,
\end{split}
\end{equation}
for any $n=0,\cdots,N-1$, with
$$Q_j^{n+1/2}:=\frac{Q_j^n+Q_j^{n+1}}{2}.$$
The above results in a $J\times J$ nonlinear system at each $n$.

The scheme is implemented by a double precision Matlab code where
the nonlinear system is solved at each step $n$ with the
\verb"fsolve" routine with initial condition there (guess) defined
by the discrete solution computed at the previous $n-1$ step.

\subsection{Soliton starter}
We first took $x_{\rm fin}:=40$, $T:=10$, $N:=200$, $J:=1200$ and
$$q(x,0)=q_0(x):=\exp(ix){\rm sech}(x),\;\;\;q(0,t)=Q(t):={\rm sech}(-t).$$
The next Figure 6.1. presents the graph of the numerical
approximation of $|q(x,t)|$ at $t=t_{fin}:=10$ as a function of
$x$ (blue line) and in red the measure of the exact solution
$$q(x,t)=\exp(ix){\rm sech}(x-t),$$
 at the same $t$. As we observe the numerical results are in good agreement with
the exact solution.
\begin{center}
\begin{figure}[!htb]\label{fig1d}
       \setlength{\abovecaptionskip}{-0.1cm}
    \includegraphics[width=8cm]{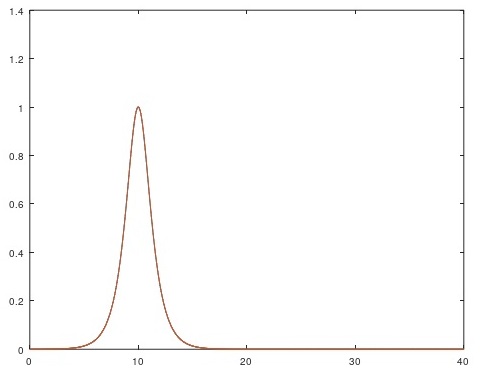}
        \caption{\textit{Numerical solution's measure versus the exact solution's measure at $t=10$.}}
\end{figure}
\end{center}

We have also computed the error between the measure of the
discrete solution and the measure of the exact solution at $t=10$
in the discrete $L^2$ norm defined by
$$E_h:=\Big{(}\sum_{j=1}^J h[|Q_j^n|-|q(x_j,t^n)|]^2\Big{)}^{\frac{1}{2}},\;\;n:=N=200,$$
and the corresponding experimental order of accuracy as presented
in the the next table.
$$
 \begin{tabular}{|c|c|c|}
   \hline
   J & $E_h$ & rate \\ \hline
   75 &1.214175838657366e+00 & - \\\hline
   150 &2.113882358596404e-01 & 2.522010374394151e+00
 \\ \hline
   300 &5.321364913254618e-02 &1.990026845331758e+00\\ \hline
   600 &1.544493946041278e-02 & 1.784662123863930e+00\\ \hline
   1200 &6.112432110480996e-03 &1.337315774488344e+00\\ \hline
 \end{tabular}
$$

Moreover, we computed the numerical error between the discrete
solution and the exact solution at $t=10-\frac{k}{2}$ in the
discrete $L^2$ norm, defined by
\begin{equation}\label{ee}
\tilde{E}_h:=\Big{(}\sum_{j=1}^J
h\Big{|}\frac{Q_j^{n}+Q_j^{n+1}}{2}-q(x_j,t^{n+1/2})\Big{|}^2\Big{)}^{\frac{1}{2}},\;\;n:=N-1=199,
\end{equation}
and the experimental order of accuracy. The numerical results are
presented in the next table.
$$
\begin{tabular}{|c|c|c|}
   \hline
   J & $\tilde{E}_h$ & rate \\ \hline
   75 &1.569997150988996e+00 & - \\ \hline
   150 &2.724195787136435e-01 & 2.526859642868486e+00
 \\\hline
   300 &5.652155291439077e-02 &2.268957384151833e+00\\ \hline
   600 &1.107594519770231e-02 & 2.351371283299980e+00\\ \hline
    \end{tabular}$$

Finally, we observed the propagation at larger distance by setting
$x_{\rm fin}:=100$ and used $T:=10$, $N:=200$, $J:=600$. Figure
6.2. presents the graph of $|q(x,t)|$ at $t=10$.
\begin{center}
\begin{figure}[!htb]\label{fig1*}
       \setlength{\abovecaptionskip}{-0.1cm}
    \includegraphics[width=8cm]{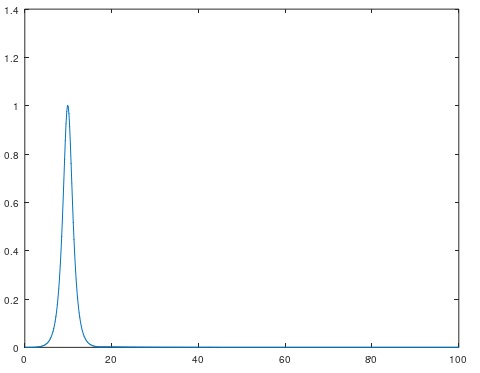}
        \caption{\textit{$|q(x,t)|$ at $t=10$.}}
\end{figure}
\end{center}
The error in the discrete $L^2$ norm, defined by \eqref{ee} at
$n=N-1=199$, was small and equal to
$$\tilde{E}_h=9.459573984096213e-02.$$
\subsection{Starters of larger measure}
We took $x_{\rm fin}:=40$, $T:=8$, $N:=200$, $J:=600$ and applied
our code for
$$q(x,0)=q_0(x):=1.8\times\exp(ix){\rm sech}(x),\;\;\;q(0,t)=Q(t):=1.8\times{\rm sech}(-t).$$
Here, note that $|q_0|$, $|Q|$ admit larger values in comparison
with the previous set of experiments and correspond to higher
initial energy. The next Figure 6.3. presents $|q(x,t)|$ computed
at times $t=0,\;t=5,\;t=8$.
\begin{figure}[!htb]
    \label{fig2d}
    \setlength{\abovecaptionskip}{-0.1cm}
     \includegraphics[width=4.05cm]{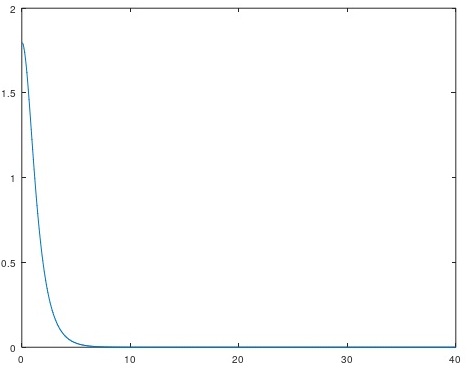}\includegraphics[width=4.05cm]{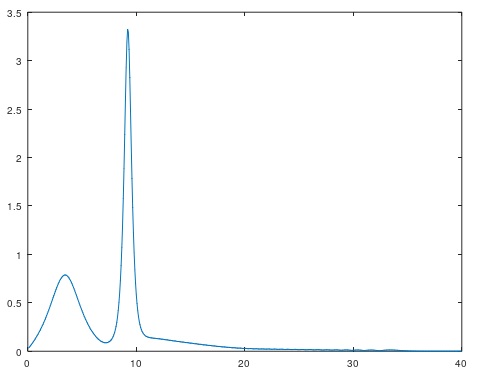}\includegraphics[width=4.05cm]{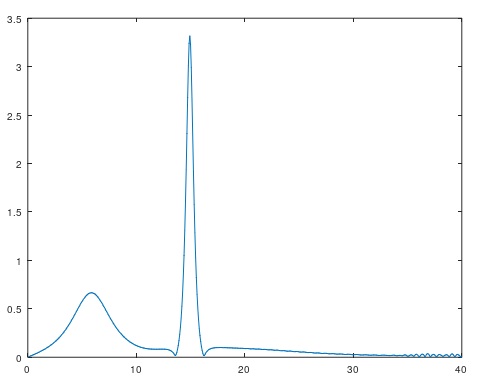}
        \caption{\textit{$|q(x,t)|$ at $t=0,5,8$.}}
\end{figure}

Increasing further the initial energy, for $x_{\rm fin}:=40$,
$T:=8$, $N:=200$, $J:=600$, we used
$$q(x,0)=q_0(x):=2.2\times\exp(ix){\rm sech}(x),\;\;\;q(0,t)=Q(t):=2.2\times{\rm sech}(-t).$$
In Figure 6.4. the set of graphs presents $|q(x,t)|$
 at times $t=0,\;t=5,\;t=8$.
\begin{figure}[!htb]
    \label{fig3d}
    \setlength{\abovecaptionskip}{-0.1cm}
     \includegraphics[width=4.2cm]{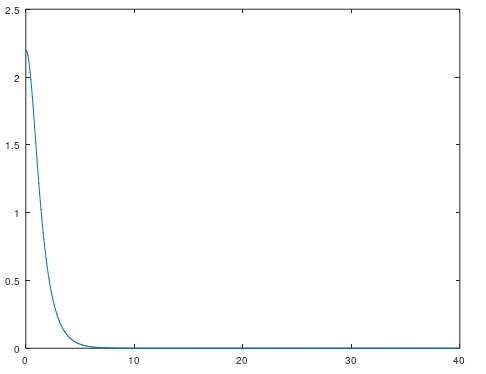}\includegraphics[width=4cm]{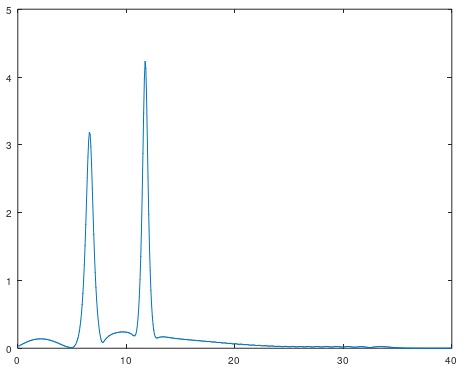}\includegraphics[width=4.2cm]{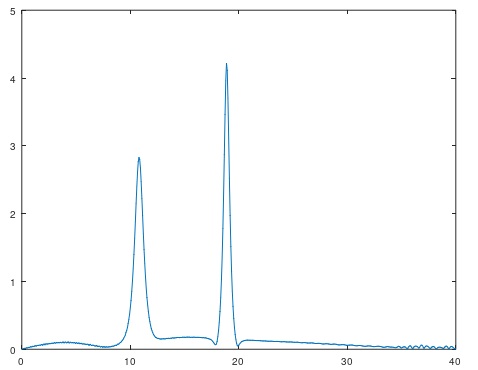}
        \caption{\textit{$|q(x,t)|$ at $t=0,5,8$.}}
\end{figure}

In both cases, as time passes, the wave seems to split in two
solitons.
\subsection{Multiple solitons}
We used $x_{\rm fin}:=50$, $T:=15$, $N:=200$, $J:=300$, and
considered
$$q(x,0)=q_0(x):=\exp(ix){\rm sech}(x),\;\;\;q(0,t)=Q(t):={\rm
sech}(-ct),$$ for $c:=\frac{1}{5}$, $c:=\frac{1}{10}$,
$c:=\frac{1}{15}$. Figure 6.5. presents $|q|$ at $t=15$ for the
previous cases.
\begin{figure}[!htb]\label{dfig5}
    \setlength{\abovecaptionskip}{-0.1cm}
     \includegraphics[width=4.0cm]{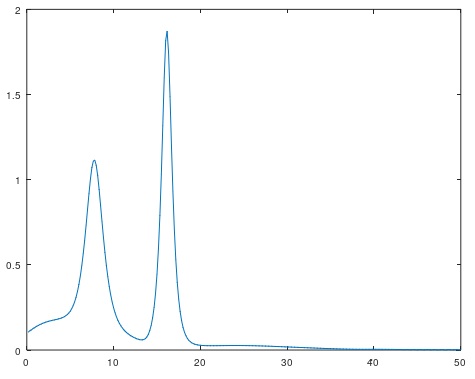}\includegraphics[width=4cm]{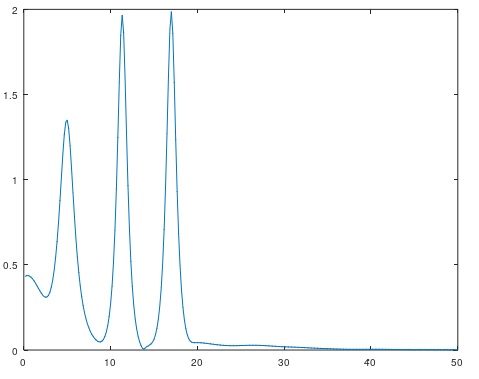}\includegraphics[width=4.2cm]{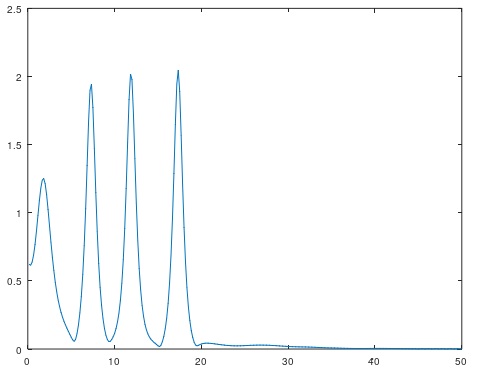}
        \caption{\textit{$|q(x,t)|$ at $t=15$ when $c:=\frac{1}{5}$, $c:=\frac{1}{10}$,
$c:=\frac{1}{15}$.}}
\end{figure}

The number of solitary waves increases as $c$ gets smaller.

\section{Conclusion. What next?}

Initial-boundary value problems for partial differential equations that admit a Lax Pair formulation $can$  give rise to  completely integrable systems that can be handled via Inverse Scattering and Riemann-Hilbert deformation methods. But there are also initial-boundary value problems for even some of the simplest Lax Pair equations where  reasonable seeming data can give rise to irregular "fractal-chaotic" behavior where the (existing) inverse methods do not apply.
Can one understand better when and why this happens?

Our generalised (vague) definition of integrability is given in reference to the existence of a Riemann-Hilbert problem with "tame" data. A definition involving an infinity of conserved quantities is obviously out of the question. But a Riemann-Hilbert problem is amenable to an asymptotic analysis which can be indicative of the "tameness" of the solution.

Can one give comprehensive sets of boundary conditions that lead to integrability and non-integrability respectively? Can one give proofs of integrability when it holds, generalising \cite{AK1} and \cite{AK2}?

Are there several degrees of non-integrability, ranging between the existence of more and less explicit asymptotic formulae respectively
to the display of completely irregular and not locally describable behavior?

Could it be that existence of a  Lax Pair (or integrability  in some sense) is related to the
existence of  bona fide fractal behavior with proper self-similarity structure?

\end{document}